\documentclass[aps,prl,twocolumn,showpacs,amsmath,amssymb,floatfix,nofootinbib]{revtex4}
\usepackage{amsmath,amssymb,amsfonts}
\usepackage{graphics,graphicx}
\usepackage{epstopdf}

\begin{document}

\title{Collective mechanics of embryogenesis: Formation of ventral furrow in {\sl Drosophila}}

\author{A. Ho\v cevar$^{1,2}$ and P. Ziherl$^{1,3}$}
\affiliation{$^1$Jo\v zef Stefan Institute, Jamova 39, SI-1000 Ljubljana, Slovenia} 
\affiliation{$^2$Department of Physics and Astronomy, University of Pennsylvania, 209 S. 33rd St., Philadelphia PA 19104-6396, USA} 
\affiliation{$^{3}$Faculty of Mathematics and Physics, University of Ljubljana, Jadranska 19, SI-1000 Ljubljana, Slovenia}

\begin{abstract}
We propose a 2D mechanical model of the ventral furrow formation in {\sl Drosophila} that is based on undifferentiated epithelial cells of identical properties whose energy resides in their membrane. Depending on the relative tensions of the apical, basal, and lateral sides, the minimal-energy states of the embryo cross-section includes circular, elliptical, biconcave, and buckled furrow shapes. We discuss the possible shape transformation consistent with reported experimental observations, arguing that generic collective mechanics may play an important role in the embryonic development in {\sl Drosophila}.  
\end{abstract}

\date{\today}


\maketitle

\begin{figure}[t!]
\includegraphics{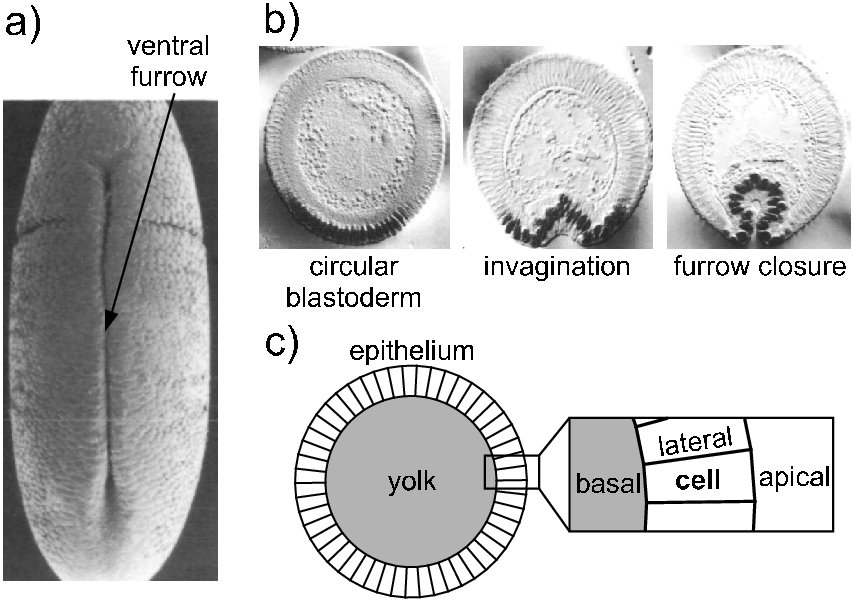}
\caption{{\sl Drosophila} ventral furrow (micrograph reproduced with permission from Ref.~\cite{Sweeton91}) is the lengthwise invagination of the embryo on ventral side (panel a). The cross-sections of the central part (reproduced with permission from Ref.~\cite{Leptin90}) show the inward buckling of the initially circular epithelial wall (panel b). In our 2D model (panel c), the cross-section consists of a ring of $N$ quadrilateral cells of area $A_c$ enclosing the yolk of area $A_y$. All apical, basal, and lateral edges are characterized by line tensions $\Gamma_a, \Gamma_b,$ and $\Gamma_l$, respectively.
}
\label{prvaslika}
\end{figure}

The single most important step in the embryonic development of all multicellular animals is gastrulation~\cite{Leptin05}. It establishes the future body plan of the organism and transforms a topologically spherical embryo into a topological torus. In the beginning of this crucial process, a part of the convex shell-like single layer of cells called the blastula folds in, and the invaginated part of the embryo develops into the digestive system~\cite{Leptin05}. The geometry of invagination differs quite considerably among the animals as does the shape of the embryo itself~\cite{Leptin05,Leptin90,Sweeton91,Hardin89}. One of the best-studied types of gastrulation is the formation of the ventral furrow in the fruit fly ({\sl Drosophila melanogaster})~\cite{Sweeton91,Davidson95,Conte08}. At the beginning of this process, the {\sl Drosophila} embryo is an elongated ellipsoidal one-cell-thick epithelium of about 6000 cells coating the yolk. During the first stage of the furrow formation, a lengthwise invagination occurs as the cells in a stripe on ventral side deform and buckle inwards~\cite{Sweeton91} (Fig.~\ref{prvaslika}a,b). This primary invagination globally alters the shape of the embryo and is one of the vital factors for determining the future head-to-tail body axis~\cite{Leptin05}. 

The role of mechanics in the early embryonic development is universally appreciated~\cite{Lecuit07}, and a range of models have been proposed to explain the formation of ventral furrow. To the best of our knowledge, all of them recognize that the invaginating cells --- the mesoderm --- are different from the rest of the embryo~\cite{Munoz07,Conte08,Conte09,Pouille08,Allena10}. By forcing the mesoderm cells into a keystone shape with a small outer (apical) face and a large inner (basal) face, the ventral section of the embryo can be driven rather naturally into an invaginated shape. This can be done by superposing active and passive deformations; the former are generated by an inherent preferred keystone shape of cells and the latter result from the elasticity of the epithelium~\cite{Munoz07,Conte08,Conte09,Allena10}. Alternatively, invagination can be reproduced by hydrodynamic motion of embryo immersed in a surrounding fluid and consisting of cells characterized by surface energy~\cite{Pouille08}. In this theory, formation of ventral furrow is triggered by an increase of the apical membrane tension of the ventral cells. These sophisticated models have been elaborated both in the simplified 2D variants describing the predominantly cylindrical central section of the embryo~\cite{Munoz07,Conte09,Pouille08} as well as in the full 3D geometry~\cite{Conte08,Allena10}, and they also included the effect of the vitelline membrane which covers the embryo~\cite{Munoz07,Pouille08,Allena10}. 

In this paper, we challenge the importance of cell differentiation as the main postulate of the existing theories of ventral furrow formation in {\sl Drosophila}. Instead we explore the primary invagination as a collective phenomenon in a system of cells of identical mechanical properties. In our 2D model of the embryo cross-section that is based on surface energy alone, each type of cell side is characterized by a specific line tension: The apical, the basal, and the lateral line tensions are all different from one another but each of them is the same in all cells. We compute the phase diagram of the model epithelium subject to constraints associated with the incompressibility of the yolk and each cell to find that it does include the invaginated shapes even though the model is devoid of any cell specificity. A more detailed analysis shows that a thick enough epithelium containing a large enough number of cells is needed for the invagination to take place.

We take our starting geometry at the time before ventral furrow formation begins and the central part of the embryo is cylindrically symmetric. Experiments show that the size of each cell along the long axis of the embryo is almost constant during the process~\cite{Sweeton91} and the cell content is incompressible. If the cells are approximated by prisms oriented lengthwise relative to the long axis of the embryo, this suggests that the area of the cell sides normal to the embryo long axis is conserved. In this case, the cross-section of the central part of the embryo captures the only variable terms in the Hamiltonian based on surface energy. 

In our 2D model, the $N$ cells in the embryo cross-section are represented by quadrilaterals arranged in a ring that encompasses a given amount of yolk (Fig.~\ref{prvaslika}c). Both the cells and the yolk are assumed incompressible~\cite{Munoz07,Pouille08} so that the area of each quadrilateral cell $A_c$ is fixed and identical for all cells, and the area of the polygonally shaped yolk $A_y$ is fixed too. Apart from incompressibility, we disregard any bulk properties of cells, yolk, and the surrounding medium. We attribute the energy of the system entirely to the three types of cell sides~\cite{Evans84,Derganc09} and the key assumption of our model is that the line tensions of lateral, basal, and apical edges are not the same. This seems plausible, because the tension on the lateral sides is determined by the cell cortex tension and on cell-cell adhesion, whereas the tensions on the apical and basal sides are controlled by the cell cortex tension and surface energy (which depends on the neighboring medium, and the yolk and the vitelline membrane are functionally dissimilar).

The energy of the model epithelium reads
\begin{equation}
W=\sum_{\rm i=1}^{N}\left(\Gamma_a L_{a}^i+\Gamma_b L_{b}^i+\frac{1}{2}\Gamma_l L_{l}^i\right).
\end{equation}
$\Gamma_a, \Gamma_b,$ and $\Gamma_l$ are the line tension of the apical, basal, and lateral edges of cells, respectively, and $L_{a}^i, L_{b}^i,$ and $L_{l}^i$ are the lengths of these respective edges in cell $i$. The sum goes over all cells and the value of each of the three line tensions is the same in all cells. Using the reduced apical and basal line tensions
\begin{equation}
\alpha=\frac{\Gamma_a}{\Gamma_l} \hspace{0.5cm}{\rm and}\hspace{0.5cm} \beta=\frac{\Gamma_b}{\Gamma_l},
\end{equation}
the total energy [Eq. (1)] can be written in dimensionless form
\begin{equation}\label{energ}
w=\alpha l_a+\beta l_b +l_l,
\end{equation}
where $l_a=\sum_iL_a^i/R_y, l_b=\sum_iL_b^i/R_y$ and $l_l=1/2\sum_iL_l^i/R_y$ are the reduced sums of all apical, basal, and lateral edge lengths, respectively. The radius of the yolk before the formation of ventral furrow $R_y=\sqrt{A_y/\pi}$ sets the length scale of the problem.

To describe the formation of ventral furrow as closely as possible, we fix the number of cells $N$ to 50 and set $A_c/A_y=1/60\approx0.0167$ as suggested by the micrographs of the embryo cross-section~\cite{Conte09}; thus the total area occupied by the epithelium is 5/6 or about 83 \% of the yolk area. The only remaining dimensionless parameters of the model are the reduced apical and basal line tensions $\alpha$ and $\beta$, respectively. By numerically minimizing the energy [Eq.~(\ref{energ})] subject to the two area constraints $A_c=const.$ and $A_y=const.$, we find the exact equilibrium shapes of the blastula. Minimization is done using the Surface Evolver package~\cite{Brakke92}.

\begin{figure*}[ht!]
\includegraphics[width=135mm]{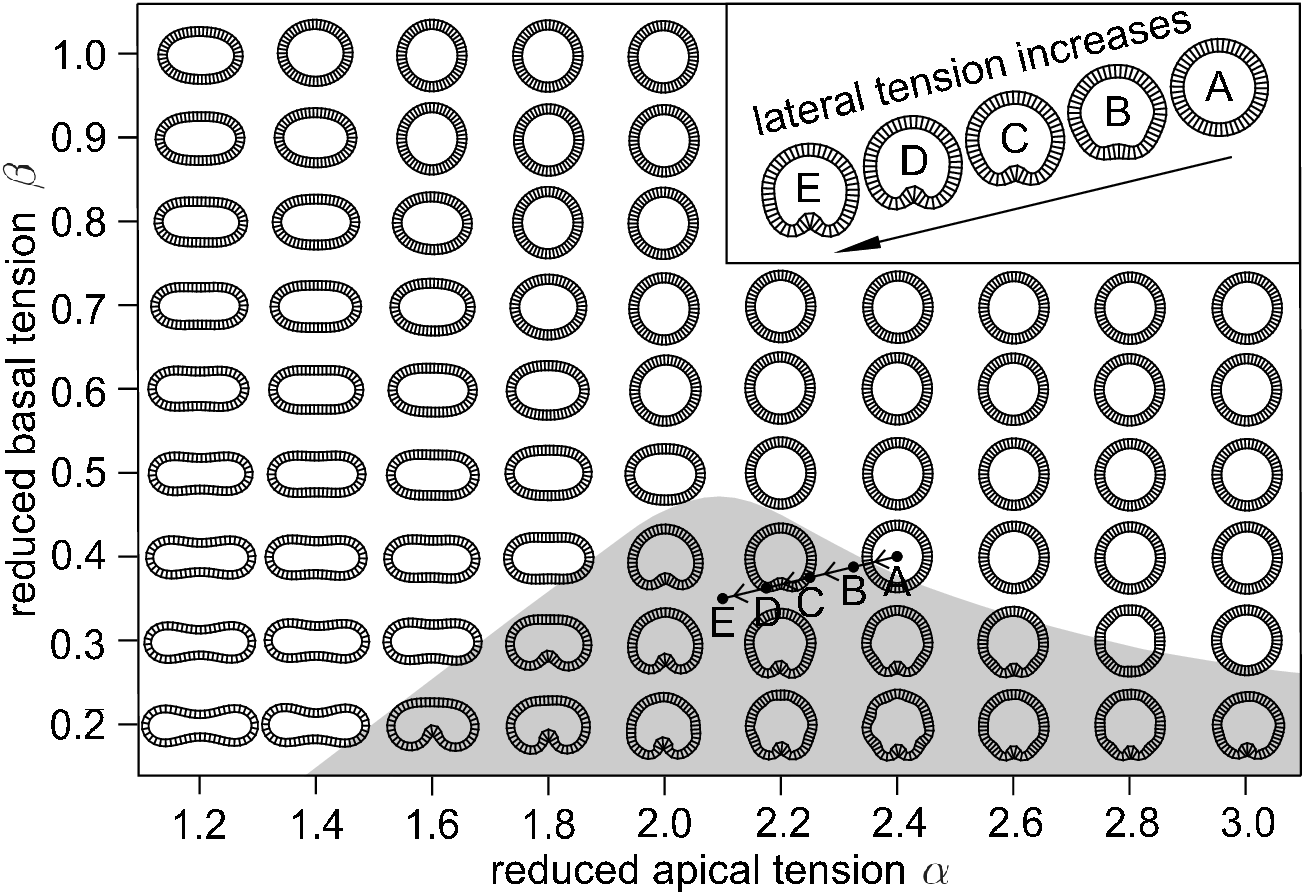}
\caption{Shape diagram of the embryo cross-section in the $(\alpha,\beta)$-plane for $N=50$ and $A_c/A_y=1/60$. The invaginated shapes qualitatively reminiscent of the {\sl Drosophila} embryo after ventral furrow formation are located in the shaded region, and those at $\alpha$ around 2.2 and $\beta$ no larger than 0.4 are quantitatively very similar to the experimental micrographs of embryo after primary invagination~\cite{Munoz07,Brodland10}. The rest of the diagram contains symmetrical elongated shapes (bottom left) and circular shapes (top right). A possible transformation of a circular shape into an invaginated one by proportionally decreasing the two reduced tensions $\alpha$ and $\beta$ (A-E) is elaborated in the inset.}
\label{slikadruga}
\end{figure*}

Our central result is the diagram of stable shapes shown in Fig.~\ref{slikadruga}. The $(\alpha,\beta)$-plane can be partitioned into three regions according to the symmetry of the shapes. At large $\alpha$ and $\beta$ the stable shapes are circular whereas at $\alpha\lesssim2$ and $\beta$ not too large the cross-section of the embryo is characterized by $C_{2h}$ symmetry. The smaller the reduced basal line tension, the more elongated the shapes in this region and the more pronounced the shallow symmetric invaginations on the long sides. The third region of the diagram (shaded part of Fig.~\ref{slikadruga}) contains buckled shapes marked by a single invagination. Some of these are globally round just like the cross-section of the {\sl Drosophila} embryo. For example, the $(\alpha=2.2, \beta=0.3)$ shape is quite similar to the experimental images of the embryo right after the primary invagination~\cite{Brodland10,Munoz07}. Many of our theoretical invaginated shapes are characterized by a very localized symmetry-breaking segment of the embryo wall.

An intriguing feature of the shape diagram is the proximity of the circular and the invaginated shapes, which suggests that a direct transformation generating the invagination is possible. Depending on the starting state, one can think of many admissible pathways but a diagonal cut involving a simultaneous decrease of both $\alpha$ and $\beta$ seems the simplest and thus the most appealing. Imagine starting with the $(\alpha=2.4, \beta=0.4)$ circular shape: Decreasing both reduced line tensions by about 13 \% will transform it into a buckled shape as illustrated in the inset of Fig.~\ref{slikadruga}, thereby reproducing the early stage of gastrulation in {\sl Drosophila}.

The reason why we choose to illustrate the ventral furrow formation by a transformation involving a proportional decrease of the two reduced line tensions is rather transparent. The simplest way of materializing this particular trajectory is to increase the lateral line tension  while keeping the apical and the basal line tension unchanged. Thus the furrow formation can be induced by a variation of a single control parameter, and the variation itself can be quite modest. A possible microscopic mechanism of increasing the lateral line tension may be a decrease of the cell-cell adhesion strength at fixed cortex tension. To obtain an even more buckled shape such as the $(\alpha=2.0, \beta=0.2)$ shape, a different mechanism where $\alpha$ and $\beta$ do not change proportionally is needed.

After having established that the proposed model predicts shapes remarkably similar of those seen in the {\sl Drosophila} embryo, we would like to understand its workings in more detail. Firstly, we vary the number of cells in the cross-section to see whether the discreteness of the model is essential. At fixed yolk and epithelium area, we halve $N$ (thus setting $A_c/A_y$ to $1/30\approx0.0333$) and recompute the shape diagram.
All other things being equal, this would halve the total energy associated with the lateral sides so it is not surprising that the shape diagram for $N=25$ can be mapped onto that for $N=50$ such that a given $(N=25,\alpha_{25},\beta_{25})$ shape corresponds to a $(N=50,\alpha_{50}=2\alpha_{25},\beta_{50}=2\beta_{25})$ shape. The only exception are the furrow shapes absent in the $N=25$ diagram, which implies that a sufficiently large number of cells is needed to stabilize the buckled shapes. This conclusion is further supported by the $N=100$ diagram, which does contain the furrow shapes and nicely agrees with the $N=50$ results after the mapping analogous to that mentioned above. In turn, it suggests that there may exist a continuum limit of our model similar to that introduced in Ref.~\cite{Derganc09}.

\begin{figure}[hb!]
\includegraphics{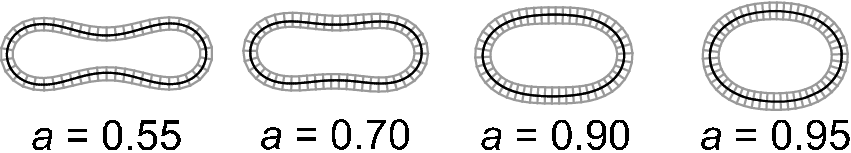}
\caption{Comparison of selected biconcave and elliptical model embryo cross-sections from Fig.~\ref{slikadruga} and the corresponding 2D vesicle contours, which are labeled by the reduced area $a$ and closely match the neutral line of the epithelium. The epithelium cross-sections from left to right correspond to ($\alpha=1.2, \beta=0.2$), ($\alpha=1.4, \beta=0.3$), ($\alpha=1.8, \beta=0.5$), and ($\alpha=2.0, \beta=0.5$). In the part of Fig.~\ref{slikadruga} occupied by the vesicle-like shapes, the reduced area is constant along lines described approximately by $\alpha/2+\beta=const.$}
\label{slikatretja}
\end{figure}

The circular, elliptical, and biconcave shapes shown in Fig.~\ref{slikadruga} closely resemble the shape of 2D lipid vesicles, infinitely thin unstretchable but flexible loops~\cite{Seifert91}. This is illustrated in more detail in Fig.~\ref{slikatretja} where four selected non-furrow shapes are replotted along with the contours of the matching 2D vesicles. 
The vesicle contours are characterized by the so-called reduced area $a=4\pi A/L^2$ (where $A$ and $L$ are the area and the perimeter of the vesicle, respectively), and they were computed using Surface Evolver~\cite{Brakke92}. The very good agreement leads us to conjecture that the buckled shapes (which do not have a 2D vesicle counterpart) must be stabilized by the finite thickness of the epithelium. 

To verify whether this is true, we varied the area of the cell while keeping the number of cells and the yolk area fixed. For $N=50$, we computed the shape diagram for $A_c/A_y=0.0333, 0.025, 0.0167, 0.0133,$ and $0.00833$. The results show that as the thickness of the epithelium is decreased, the region of the phase space where the buckled shapes are stable gradually shrinks. For $N=50$, the buckled shapes are present only in model embryo cross-sections with $A_c/A_y\geq0.0133$. In this sense, gastrulation in the elongated {\sl Drosophila} embryo is very different from that in the spherical sea urchin embryo which can be qualitatively reproduced by a theory based on the elasticity of a thin fluid shell~\cite{Bozic06}.

There are a few features of the ventral furrow formation that are not captured by our model, the most important one being the approach and the closing of the furrow (final stage in Fig.~\ref{prvaslika}b). These processes can be phenomenologically included by attraction and adhesion of apical faces of cells on the opposite sides of the furrow. In addition, the scenario proposed above involves equilibrium states of the epithelium and disregards the hydrodynamics of the furrow formation, which could be introduced like in Ref.~\cite{Pouille08}. Elaborating our model to account for these aspects of the process seems feasible, but simplifying it in any way without compromising its predictions is likely much more difficult.

The main conclusion of this work is that from the mechanical perspective, neither cell differentiation~\cite{Groshans00} nor the ensuing inhomogeneous active or passive deformations are needed for the primary invagination in the ventral furrow formation. This is consistent with the attenuated furrow seen in mutants with suppressed mesoderm differentiation, which involves fewer cells and is not as deep as in normal embryos~\cite{Ip94}. Moreover, our results show that the buckled yet overall round shape is stable in absence of the vitelline membrane, suggesting that the ventral furrow in {\sl Drosophila} can be shaped by the generic, spatially uniform, and collective mechanism proposed here. Its role relative to the localized morphogenetic processes driven by cell differentiation is hard to assess, but the two are probably active simultaneously. If embryogenesis were engineered, differentiation would probably harness and steer the thrust provided by the buckling of the thick epithelium.

We thank G. Belu\v si\v c, C.-P. Heisenberg, A. Jacinto, R. D. Kamien, S. Svetina, and A. \v Siber for helpful discussions. A. H. acknowledges the hospitality of the Department of Physics and Astronomy, University of Pennsylvania, where a part of this study was done. This work was supported by Slovenian Research Agency through Grant No. P1-0055, by National Science Foundation through Grant No. DMR05-47230, and by a grant from Slovene Human Resources Development and Scholarship Fund.

\bigskip

\end{document}